\documentclass[]{spie}  

\usepackage{amsmath,amsfonts,amssymb}
\usepackage{graphicx}
\usepackage{booktabs}
\usepackage[colorlinks=true, allcolors=blue]{hyperref}

\title{Towards High-Throughput Visible Photonic Lanterns for the EMARCOT Project}

\author[a]{Abani Shankar Nayak}
\author[a]{Julius Goehring}
\author[b]{Marina Centenera-Merino}
\author[b]{Pedro José Amado-González}
\author[b]{David Pérez-Medialdea}
\author[b]{Mabel Ruiz-Lopez}
\author[b]{Francisco J. Pozuelos}
\author[b]{Miguel Andrés Sánchez Carrasco}
\author[b]{José Luis Ortiz Moreno}
\author[b]{Jesús Aceituno-Castro}
\author[c]{Stefan Cikota}
\author[c]{Javier Flores}
\author[d]{Christopher Betters}
\author[d]{Andrew Ross-Adams}
\author[d]{Sergio Leon-Saval}
\author[a]{Martin M. Roth}
\author[a]{Kalaga Madhav}

\affil[a]{Leibniz-Institut für Astrophysik Potsdam (AIP), An der Sternwarte 16, 14482 Potsdam, Germany}
\affil[b]{Instituto de Astrofísica de Andalucía, Consejo Superior de Investigaciones Científicas, Glorieta de la Astronomía, s/n, 18008, Granada, Spain}
\affil[c]{Centro Astronómico Hispano en Andalucía (CAHA), Observatorio de Calar Alto, Sierra de los Filabres, 04550 Gérgal, Almería, Spain}
\affil[d]{Institute of Photonics and Optical Science, School of Physics, University of Sydney, Sydney, NSW 2006, Australia}

\authorinfo{Further author information: (Send correspondence to A.S.N.)\\A.S.N.: E-mail: anayak@aip.de, Telephone: +49 331 7499 673}

\begin{document} 
\maketitle

\begin{abstract}
Photonic EMARCOT is an innovative project involving Spanish, German and Australian research institutes that aims to integrate multiple Optical Tube Assemblies (OTAs) using photonic lanterns. The "Pathfinder" prototype, featuring seven OTAs with a 1.1-meter effective aperture, will feed a spectrograph at the Calar Alto Observatory, with first light expected in 2026. We report the fabrication and experimental evaluation of a custom $7\times1$ multi-mode photonic lantern (MMPL) developed for this framework, featuring seven $25\,\mu m$ core multi-mode inputs merging into a single $50\,\mu m$ core multi-mode output optimized for the visible wavelength range (400–700 nm). Optical characterization centered at 600 nm reveals exceptional channel-to-channel uniformity, with statistical variations close to zero across both bare-fiber and connectorized MMPL configurations. However, the total baseline throughput of this initial device was limited to below $4\%$. From the refractive index studies, this low throughput is attributed to severe refractive index mismatch between the internal fiber cladding geometry and the structural capillary, which suppresses total internal reflection during the tapering transition. This work establishes an important diagnostic baseline that highlights the necessary fabrication tolerances needed to improve future high-throughput manufacturing processes for precision radial-velocity astronomy.
\end{abstract}


\section{INTRODUCTION}
\label{sec:intro}  
The Photonic European Multi-Array of Combined Telescopes (EMARCOT) represents an innovative, scalable, and highly cost-effective solution for modern astronomical facility design. Developed through a joint collaboration among Spanish, German, and Australian research institutes, EMARCOT is engineered to deliver high-resolution spectroscopy and wide-field, high-dynamic-range imaging at sub-arcsecond resolution \cite{EMARCOT}, with first light expected in 2026. By synthesizing a large effective collecting aperture from an array of smaller optical tube assemblies (OTAs), the architecture substantially lowers capital costs relative to monolithic telescopes of equivalent aperture class. As a critical step toward validating this concept on-sky, a 7-telescope prototype array -- the MARCOT Pathfinder -- has been successfully constructed and deployed at the Calar Alto Observatory \cite{roth_2022}. 

A key technological enabler for combining multiple identical OTAs to simulate a giant aperture framework is the photonic lantern (PL) \cite{Birks:12, Leon-Saval:10}. Acting as a crucial link between classical observational astronomy and guided-wave optics, photonic lanterns have significantly advanced the field of astrophotonics by allowing for the precise manipulation of starlight within integrated systems \cite{Bland-Hawthorn:09}. Over the past two decades, extensive research has transformed these components from laboratory proofs of concept into reliable on-sky instruments \cite{Jovanovic_2023}.

A classical photonic lantern is a guided-wave device that transitions smoothly from a multi-mode (MM) waveguide entrance to a carefully arranged bundle of isolated single-mode (SM) cores \cite{Noordegraaf:09}. At the multi-mode interface, complex or aberrated light from a telescope's focal plane is captured. As the light travels through the downstream longitudinal taper, its spatial profile is gradually transformed into individual single-mode outputs \cite{Leon-Saval:10}. If the geometric transition is sufficiently gradual and the number of independent single-mode ports equals or exceeds the total number of spatial modes supported at the multi-mode boundary, this conversion occurs essentially without loss, preserving the fundamental conservation theorem \cite{Birks:15}.

In astronomical instrumentation, photonic lanterns are commonly used to interface with adaptive optics (AO) systems. These lanterns uniquely combine the high injection throughput of multi-mode fibers with the ideal diffraction-limited stability of single-mode systems \cite{Spaleniak:13}. Beyond simple routing \cite{Harris_2018, Harris_2020}, single-mode photonic lanterns (SMPLs) have gained widespread application as platforms for fiber Bragg grating OH-suppression filters, which isolate near-infrared sky lines \cite{LeonSaval_2013} and are currently being explored for the development of compact spectrographs \cite{LeonSaval_2013, Rahman_2026}. 

However, while traditional single-mode output lanterns excel under diffraction-limited or highly optimized AO conditions, their coupling performance degrades significantly when subjected to uncorrected, seeing-limited wavefronts. Both theoretical models \cite{Birks:15} and empirical studies \cite{Leon-Saval:17} demonstrate that a photonic lantern constructed with few-mode or multi-mode legs (which support a higher modal capacity per channel) captures substantially more optical power from atmospherically distorted beams than an equivalent lantern restricted to single-mode outputs.

For a modular, seeing-limited telescope array like EMARCOT, using a multi-mode to multi-mode photonic lantern (MMPL) configuration is very beneficial. This setup increases the spatial mode acceptance window at the input, allowing the device to handle the larger etendue of a turbulent, seeing-limited point spread function (PSF) while still maintaining high coupling efficiencies that are comparable to standard multi-mode fiber performance. Furthermore, aligned with CAHA and European research priorities, this modular system serves as an open invitation for new partners to contribute to EMARCOT's expanding instrumentation and collaborative framework (see Amado González et al. \cite{pedro_spie_2026} in these proceedings for an overview of the scientific and collaborative framework). 

In this proceeding, we present the development, fabrication, and experimental evaluation of a custom 7$\times$1 MMPL tailored specifically for the MARCOT Pathfinder system. The geometric design and operational concept of the $7\times1$ MMPL presented here are based on a proprietary light-guiding architecture developed by our team~\cite{Davenport_2023}. While the initial patent framework established the device performance via numerical simulations, this paper focuses on translating those simulated designs into actual experimental settings. Our MMPL device comprises seven input channels fabricated from commercial $25 \, \mu m$ core multi-mode fiber (Thorlabs FG025LJA), which smoothly merge into a single $50 \, \mu m$ core multi-mode output fiber (Thorlabs FG050LGA). This specific configuration allows each individual input arm to be directly coupled to one of the seven independent OTAs comprising the Calar Alto Pathfinder MARCOT array (see Sánchez Carrasco et al. \cite{miguel_2026} in these proceedings for a comprehensive facility overview). 

Our development is occurring alongside parallel efforts within the EMARCOT collaboration, which aims to further optimize manufacturing workflows to maximize the device throughput of MMPL (see Centenera-Merino et al. \cite{marina_2026}). This also includes advanced numerical modeling and quantum optical studies that utilize similar photonic lantern geometries (refer to Centenera-Merino et al. \cite{marina_spie_2026} in these proceedings). In the context of these ongoing initiatives, this work primarily focuses on rigorously analyzing the low throughput of our preliminary MMPL, which demonstrated a baseline transmission efficiency to below $4 \%$. This evaluation aims to identify the internal loss mechanisms in our current MMPL, establishing a baseline to inform future fabrication adjustments and tolerance improvements.

The remainder of this paper details the fabrication methodology of the MMPL, outlines the optical characterization setups utilized to benchmark performance, and evaluates critical metrics -- including relative transmission efficiencies, standard deviation, and channel-to-channel uniformity -- across both the raw bare-fiber and connectorized MMPL.

\section{Fabrication}
\label{sec:fabrication}

The fabrication of the $7 \times 1$ MMPL starts with the preparation of all necessary fiber components. In this work, seven MM input fibers with a core diameter of $25\, \mu m$ and one output MM fiber with a $50\, \mu m$ core were used. All fibers were cleaned with isopropanol alcohol (IPA) to remove dust and surface residues. The input fibers were carefully stripped in designated sections and then inserted into a custom-made glass capillary produced by the Leibniz Institute of Photonic Technology (IPHT) in Jena, Germany. The capillary was cleaned with isopropyl alcohol (IPA). A vacuum pump was connected to the capillary to remove any remaining IPA, as any residual solvent could interfere with the tapering process that followed. Finally, the loaded capillary was mounted in the Vytran GPX-3000 tapering system. During tapering, the capillary and the enclosed fibers are locally softened and drawn into a unified structure -- the photonic lantern -- where the individual fibers form the guiding cores and the capillary becomes the surrounding cladding. The taper consists of three regions: a downtaper, where the inner diameter decreases from $410\,\mu m$ to $50\,\mu m$; a uniform waist section of $50\,\mu m$; and an uptaper, where the inner diameter increases back to $410\,\mu m$. Following tapering, the MMPL was transferred to the \textit{Vytran LDC-400} cleaver. The device first applied a controlled tension to the PL. A diamond blade then repeatedly contacted the PL at the waist section until a crack initiated and propagated cleanly through the structure, producing a flat, high-quality end facet suitable for splicing. To complete the device, the cleaved PL was spliced to the designated output fiber using the \textit{Vytran GPX-3000} equipped with the appropriate filament.

\section{Characterization}
\label{sec:characterization}
To evaluate the optical performance and channel uniformity of the fabricated $7\times1$ multi-mode photonic lantern (MMPL), a systematic two-phase characterization campaign was conducted. The main objective was to quantify the efficiency and uniformity of the device within its intended operating band, centered around 600 nm. Initially, characterization was performed on the raw, glued "bare" MMPL to establish an intrinsic performance baseline and identify losses associated with the manufacturing processes. As described in Sec.~\ref{sec:fabrication}, this process primarily involves tapering the fiber bundle, cleaving the tapered end, and splicing it with a standard multi-mode fiber. Following this initial stage, the lantern was equipped with standard fiber patch cords to meet the "plug-and-play" requirements for the EMARCOT facility. This section outlines the distinct experimental configurations used for both the bare and packaged MMPL, presents the raw and normalized spectral throughput results, and analyzes the physics that govern the system's overall efficiency.

\subsection{Bare Photonic Lantern}
\label{sec:bare_photonic_lantern}

\begin{figure}
\begin{center}
\begin{tabular}{c} 
\includegraphics[width = 0.9\textwidth]{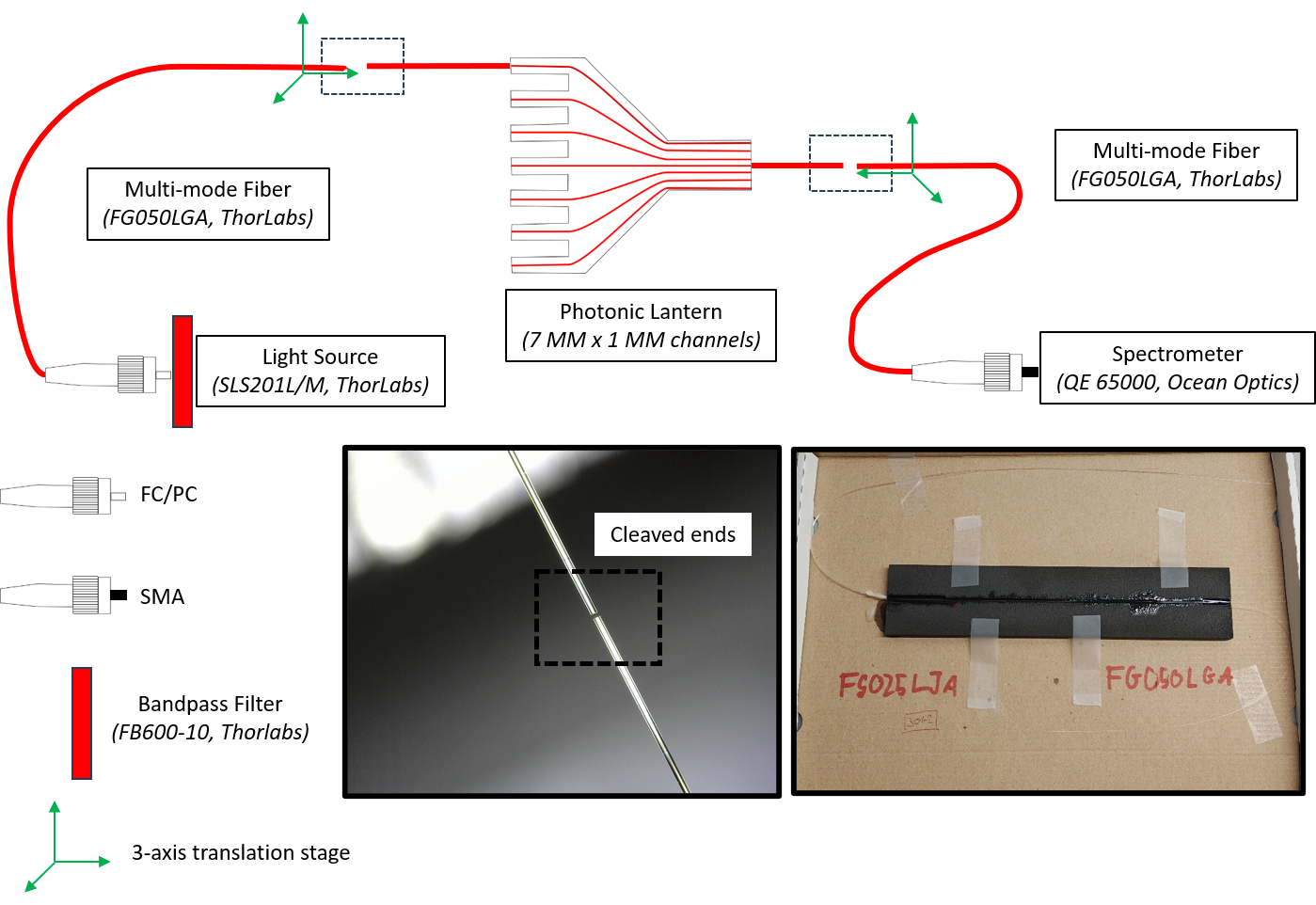}
\end{tabular}
\end{center}
\caption[] 
{\label{fig:schematic_1} A schematic of the experimental characterization setup illustrating the selection of components and the types of connections used. Active alignment between the fiber ends and the photonic lantern is achieved with two 3-axis translation stages. The optical output from the multi-mode end of the lantern is collected and routed through a standard multi-mode fiber to a spectrometer (QE 65000, Ocean Optics). The injection and collection strategy for the photonic lantern follows the EMARCOT methodology. \textit{Left Inset}: A magnified view showing the cleaved bare ends of the fiber and the PL during alignment. A clean, flat cleave is essential to minimize coupling losses in the setup. \textit{Right Inset}: Photograph of the $7 \times 1$ MMPL, which is directly glued in the groove of the custom-made 3D-printed plate that provides critical mechanical protection for the fragile tapered and spliced regions of the bare lantern.}
\end{figure}

The experimental setup designed to evaluate the transmission behavior of the MMPL is illustrated in Fig. \ref{fig:schematic_1}. A broadband light source (Thorlabs, SLS201L/M) filtered at $600 \pm 5 \, nm$ provides illumination, which is introduced into a standard multi-mode fiber patch cord (Thorlabs, FG050LGA) through a standard FC/PC connector interface. To achieve precise mode excitation, the end of the input patch cord is stripped and cleaved. This cleaved bare fiber end (as shown in the inset of Fig.~\ref{fig:schematic_1}) is mounted on a high-precision 3-axis translation stage, allowing for active alignment with an individual multi-mode (MM) input channel of a $7 \times 1$ photonic lantern. The lantern converts these seven discrete MM input channels into a single, centralized multi-mode output channel.

The optical output from the lantern is collected using a multi-mode fiber patch cord identical to that at the output end of the MMPL. This collection fiber is mounted on a separate 3-axis translation stage to optimize the coupling efficiency from the lantern's output port. The collected signal is then directed into a spectrometer (Ocean Optics, QE 65000) for precise spectral analysis. The integration time of the spectrometer was set to $16 \text{ms}$ for all channels. 

The bare lantern structure requires special protection due to its sensitivity to environmental factors and handling stresses. To address this, a custom 3D-printed plate with a groove has been designed, as illustrated in the inset photograph of Fig.~\ref{fig:schematic_1}. First, the lantern is positioned in the groove, and an adhesive is applied throughout the entire groove, which holds the fragile tapered structure and the spliced region of the lantern. The adhesive is allowed to cure for a full day to ensure complete hardening and maximum bonding strength. This process effectively encapsulates both the transition taper of the photonic lantern and the surrounding fiber splice junctions. By shielding these mechanically vulnerable areas from bending, twisting, and exposure to atmospheric contaminants, the 3D-printed plate enhances structural integrity and ensures performance stability during the alignment and characterization steps conducted in the laboratory. It should be noted that previous experiments, which are not detailed here, showed that the performance of the MMPL is minimally impacted by gluing.

\subsubsection{Results}
\label{sec:bare_pl_results}
\begin{figure}
   \begin{center}
   \begin{tabular}{cc} 
   \includegraphics[width = 0.45\textwidth]{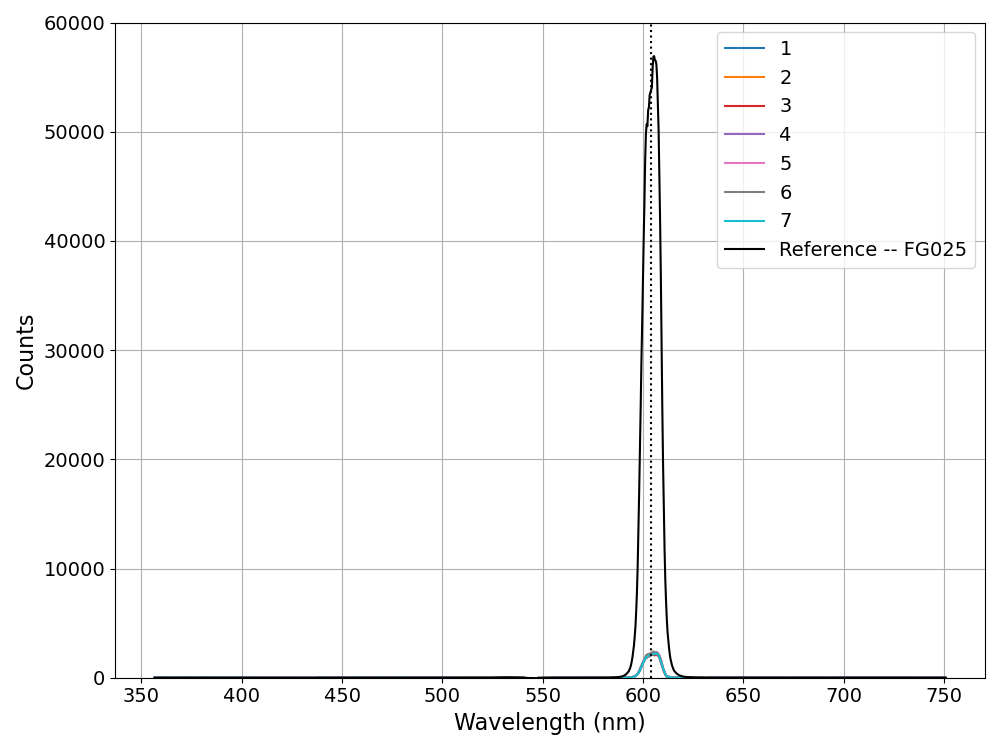} &
   \includegraphics[width = 0.45\textwidth]{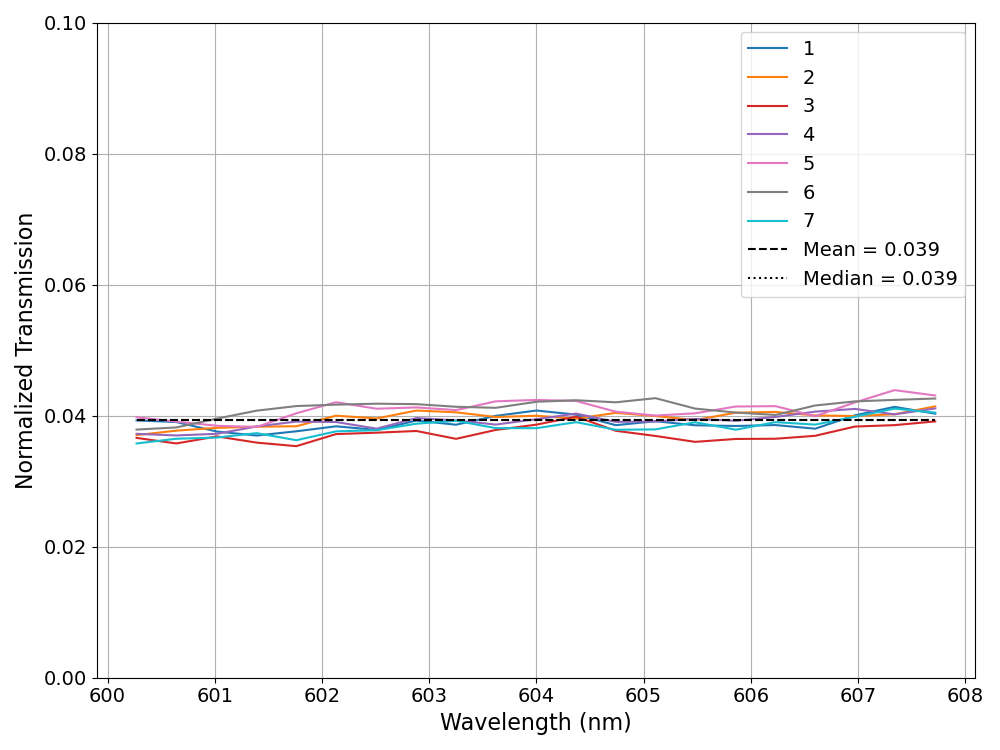}\\
   \small(a) & \small(b) 
   \end{tabular}
   \end{center}
   \caption[] 
   {\label{fig:bare_pl_results} Experimental spectra of the $7\times1$ bare MMPL. (a) The dark-corrected raw spectrum, displaying the filtered source throughput centered near 604 nm (as shown in black dotted) across all channels compared to a reference multi-mode fiber (Thorlabs FG025LJA, $25\, \mu m$ core) of length $1-\text{m}$. (b) The calculated normalized transmission within the region of interest (600 nm to 608 nm) that has a flat spectra for all channels . Horizontal dashed and dotted lines indicate the mean and median transmission values across all channels and bands, respectively, demonstrating uniform performance at approximately 0.039 (3.9 $\%$) relative to the reference fiber.}
   \end{figure}

The setup shown in Fig. \ref{fig:schematic_1} was used to evaluate the spectral performance and channel uniformity of the MMPL, with the results summarized in Figure 2. The raw, dark-corrected spectrum obtained from the spectrometer is displayed in Fig. \ref{fig:bare_pl_results}. Individual spectrum of the 7 lantern channels are also plotted alongside a baseline spectrum taken from a $25\text{ }\mu\text{m}$ core multi-mode reference fiber (Thorlabs, FG025LJA) of length $1-\text{m}$. The bandpass filter used in the setup was expected to center the spectrum at 600 nm, but it shifted slightly to 604 nm, which is within the filter's manufacturing tolerance. From repeated trials, the filter showed a stable center wavelength of 604 nm, which is suitable for this experiment.

To evaluate transmission efficiency, data from all channels were extracted within a region of interest spanning 600 nm to 608 nm and normalized to the reference fiber spectrum shown in Fig. \ref{fig:bare_pl_results}b. In this case, the total efficiency ($\eta$) = mean transmission ($\mu$) = $\frac{1}{7} \sum_{}^{7} \eta_i$ = 0.039 ($3.9\%$). All relevant parameters and metrics are presented in Tab.~\ref{tab:pl_parameters}. 

\begin{figure}
\begin{center}
\begin{tabular}{c} 
\includegraphics[width = 0.5\textwidth]{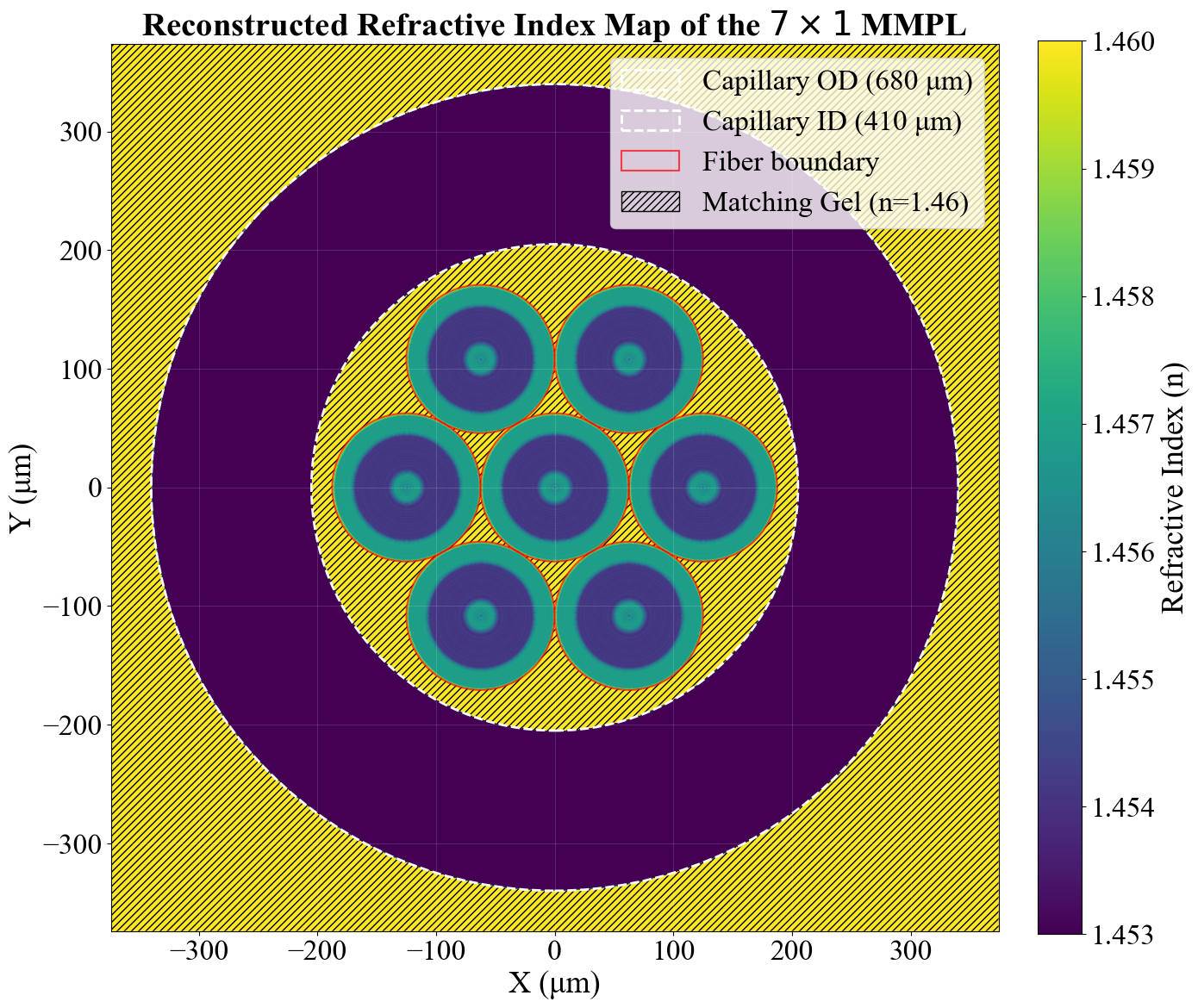}
\end{tabular}
\end{center}
\caption[] 
{\label{fig:refractive index_variation} Reconstructed cross-sectional refractive index profile map of the $7\times1$ MMPL inside the glass capillary, replotted for clarity from measured fiber and capillary parameters. The profile highlights the spatial distribution of the seven bundled FG025LJA fibers embedded in an index-matching gel ($n = 1.46$). As seen in the colorbar, the individual fibers exhibit a distinct double-cladding geometry: a high-index central core, an inner low-index cladding, and an outer high-index cladding. Because the refractive index of the inner cladding layer is nearly identical to that of the surrounding capillary material, the index contrast ($\Delta n$) required for effective guiding is severely degraded during the tapering process. This poor waveguide confinement suppresses efficient total internal reflection (TIR) in the lantern's transition zone, explaining the high radiative losses and the resulting $3.9\%$ throughput efficiency of the $7\times1$ MMPL.}
\end{figure}

To investigate the physical mechanisms that limit the throughput of the device, we analyzed the cross-sectional refractive index profile of the fabricated $7\times1$ MMPL using a refractive index profilometer. The calculated values were replotted for clarity and better understanding, as shown in Fig.~\ref{fig:refractive index_variation}. This map consists of seven multi-mode fibers (Thorlabs FG025LJA) that are positioned inside a structural glass capillary and an index-matching gel ($n = 1.46$). In an ideal adiabatic photonic lantern transition, the individual fiber cores gradually lose their guiding capability as the assembly is fused and elongated. Light leaks out of the shrinking cores and is guided by the fused fiber claddings, which together function as a new composite multi-mode "core." Meanwhile, the low-index glass capillary serves as the new common "cladding" layer~\cite{Davenport:21}.

However, the profilometer measurement indicates a complex double-cladding geometry that is intrinsic to the FG025LJA fiber structure, which disrupts the guiding mechanism. Each fiber consists of a high-index central core surrounded by an inner low-index cladding, which is then encased in an outer high-index cladding layer. As shown in the index map (Fig. ~\ref{fig:refractive index_variation}), the refractive index of the inner cladding layer closely matches that of the capillary material. During the tapering process, as the light transitions into a cladding-dominated regime, the minimal refractive index contrast ($\Delta n \approx 0$) hampers the formation of a solid waveguide boundary at the capillary interface. Without a significant index step to support total internal reflection (TIR), the optical field cannot be effectively confined within the fused multi-mode core. Consequently, a substantial portion of the optical power either escapes into unguided cladding modes or radiates out of the capillary walls entirely. This significant waveguiding breakdown leads to high baseline insertion losses and directly contributes to the low measured transmission efficiency of $3.9 \%$.

Though the normalized transmission of the PL across the characterization window averaged $3.9 \%$ relative to the $25\, \mu m$ core reference fiber. However, as indicated in Tab. ~\ref{tab:pl_parameters}, a coefficient of variation (CV) of $4\%$ across channels demonstrates excellent uniformity within the MMPL, meaning each of the seven input fibers experiences the same transformation matrix as it transitions into the multi-mode core.

\subsection{Connectorized Photonic Lantern}
\label{sec:connectorized_photonic_lantern}

\begin{figure}
\begin{center}
\begin{tabular}{c} 
\includegraphics[width = 0.9\textwidth]{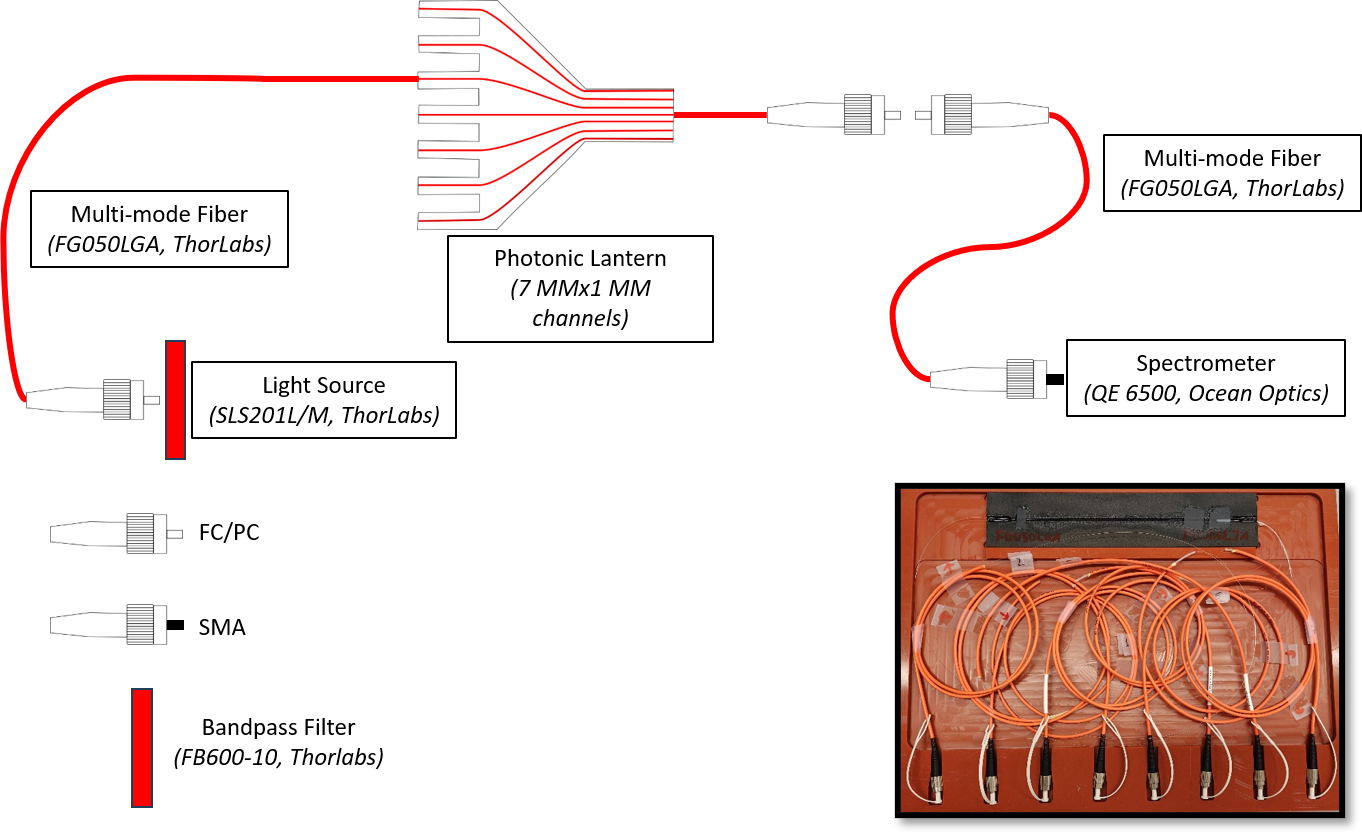}
\end{tabular}
\end{center}
\caption[] 
{\label{fig:schematic_2} A Schematic of the setup for characterizing the fully connectorized PL. \textit{Right Inset}: Photograph of the fully connectorized $7\times1$ MMPL device. To facilitate secure packaging and shipping, the MMPL is mounted on a milled plastic baseplate (orange-colored), which provides necessary mechanical stabilization and support, and protects the fragile fiber components from structural stress.}
\end{figure}

The experimental setup used to characterize the fully connectorized MMPL is shown in Fig.~\ref{fig:schematic_2}. This configuration incorporates nearly identical components to those utilized in the bare-fiber setup (Fig.~\ref{fig:schematic_1}); however, the critical distinction is that the MMPL is now completely connectorized with dedicated patch cords. This design enables efficient "plug-and-play" compatibility with standard laboratory components as well as existing optical interfaces at telescope sites. It is to be noted that the integration time of the spectrometer was set to $12 \text{ms}$, which differs from the one used in the bare setup shown in Fig. \ref{fig:schematic_1}. 

During characterization, the input patch cord connects directly to one of the seven integrated multi-mode input channels of the packaged photonic lantern via a standard FC/PC interface. These seven channels subsequently converge into a single connectorized multi-mode output channel with a $50 \,\mu m$ core diameter. In the context of the EMARCOT framework, each Optical Tube Assembly (OTA) will feed light into a combiner Multi-fiber Termination Push-on (MTP) assembly. A matching splitter MTP interface then routes this light directly into the seven connectorized input channels of the MMPL. Ultimately, the combined output channel of the MMPL delivers the signal to a downstream spectrograph.

A photograph of the fully assembled, connectorized MMPL is featured in the inset of Fig.~\ref{fig:schematic_2}. To secure the system, a custom-milled orange plastic baseplate was fabricated to enhance the structural integrity of the MMPL assembly. This substrate serves as a rigid, ruggedized platform designed for secure device packaging, safe laboratory handling, and commercial transit to the EMARCOT facility site.

\subsubsection{Results}
\label{sec:connectorized_pl_results}
\begin{figure}
   \begin{center}
   \begin{tabular}{cc} 
   \includegraphics[width = 0.45\textwidth]{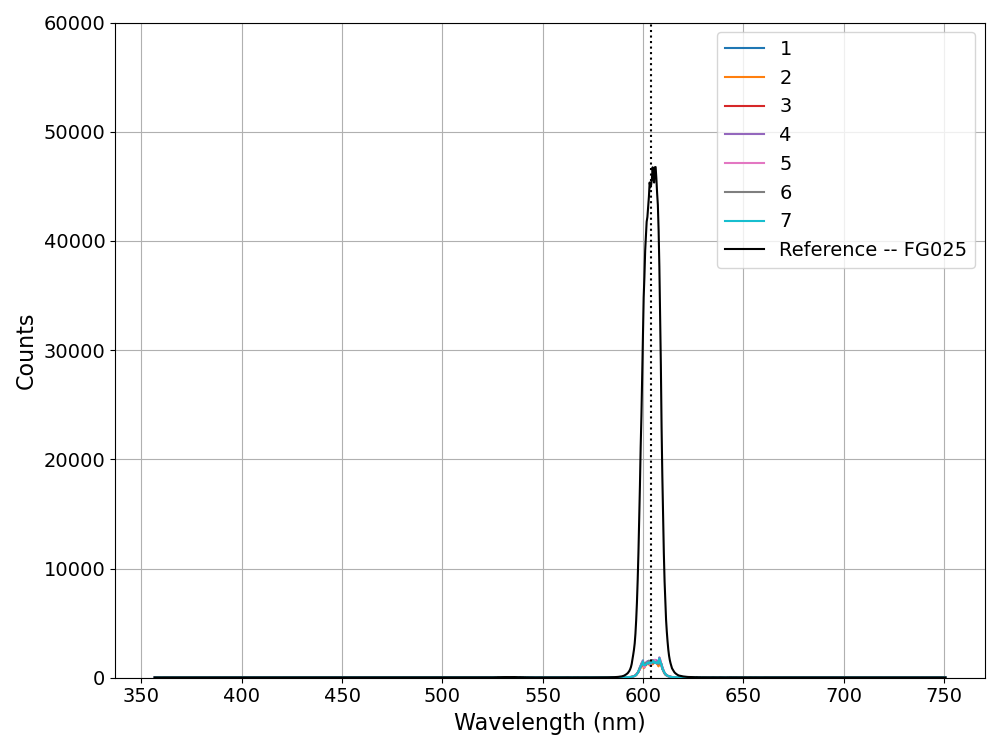} &
   \includegraphics[width = 0.45\textwidth]{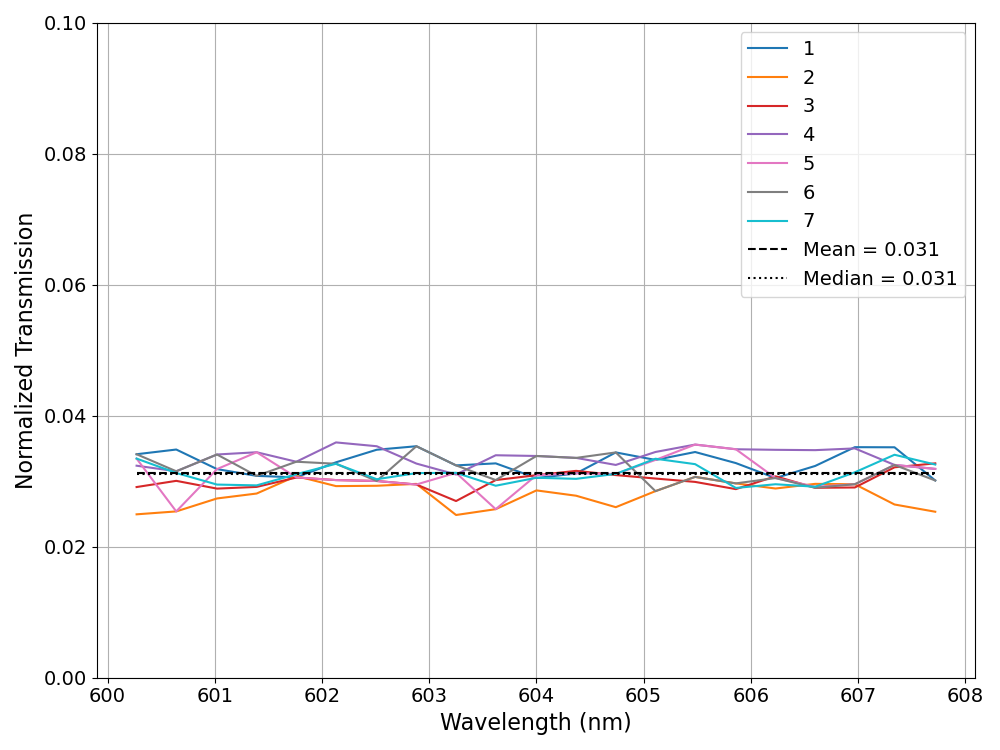}\\
   \small(a) & \small(b) 
   \end{tabular}
   \end{center}
   \caption[] 
   {\label{fig:connectorized_pl_results} Experimental spectra of the $7\times1$ connectorized MMPL. (a) The dark-corrected raw spectrum showing the filtered source throughput centered near 604~nm (as shown in black dotted) across all channels compared to a reference multi-mode fiber (Thorlabs FG025LJA, $25\, \mu m$ core) of length $2-\text{m}$. (b) The calculated normalized transmission within the region of interest (600~nm to 608~nm) that has a relatively flat spectra for all channels even after connectorization. Horizontal dashed and dotted lines indicate the mean and median transmission values, respectively, demonstrating a stable baseline throughput of approximately 0.031 (3.1 $\%$)
   relative to the reference fiber with high channel-to-channel uniformity.}
   \end{figure}

The spectral performance and channel uniformity of the fully connectorized MMPL were evaluated to assess the device's behavior after it was connectorized with patch cords. Figure \ref{fig:connectorized_pl_results}(a) shows the dark-corrected raw spectrum captured by the spectrometer. The filtered broadband illumination is centered around 604 nm, providing a spectral testing window identical to that used during the bare fiber characterization phase. The individual throughput for each of the seven connectorized channels is plotted alongside the reference spectrum obtained from a 25 µm core multi-mode reference fiber (Thorlabs, FG025LJA) with a length of 2 meters.

To isolate the device's performance, the transmission spectrum was normalized against the 2-meter reference fiber across the region of interest, from 600 nm to 608 nm, as shown in Figure \ref{fig:connectorized_pl_results}(b). The mean and median normalized transmission across all channels and bands were measured at 0.031 ($3.1 \%$). This represents a decrease of approximately $20\%$ from the $3.9\%$ throughput observed in the bare fiber setup. This additional loss may be attributed to splice losses when connectorizing the bare lanterns with the patch cords, as well as insertion losses within the mechanical FC/PC mating sleeves used in the setup. 

Despite the reduction in absolute throughput, the device maintains excellent spatial and spectral uniformity as presented in Tab.~\ref{tab:pl_parameters}. The relative coefficient of variation (CV) is approximately $8\%$ across the entire operating window. This indicates that each of the seven input fibers undergoes an identical modal transformation upon entering the multi-mode core, even after fully connectorizing the bare MMPL with patch cords. Additionally, the custom-milled plastic baseplate provides mechanical robustness by offering a "plug-and-play" architecture that eliminates the need for laboratory translation stages, making the packaged MMPL highly suitable for direct integration into downstream spectrographs at the EMARCOT facility telescope sites.

\begin{table}
\centering
\caption{Comparative performance parameters and statistical metrics for the $7\times1$ MMPL in bare and fully connectorized configurations.}
\label{tab:pl_parameters}
\resizebox{0.95\textwidth}{!}{
\begin{tabular}{|l|c|c|}
\toprule
\textbf{Parameter / Metric} & \textbf{Bare MMPL Configuration} & \textbf{Connectorized MMPL Configuration} \\ 
\midrule
\multicolumn{3}{|l|}{Individual Channel Efficiencies ($\eta_{i}$)} \\
1 & 0.0389 & 0.0328 \\
2 & 0.0396 & 0.0279 \\
3 & 0.0372 & 0.0300 \\
4 & 0.0391 & 0.0337 \\
5 & 0.0409 & 0.0310 \\
6 & 0.0411 & 0.0317 \\
7 & 0.0382 & 0.0310 \\
\addlinespace
\multicolumn{3}{|l|}{Overall System Metrics}\\
Total Efficiency ($\eta$) & 0.0393 (3.93\%) & 0.0312 (3.12\%) \\
Mean Transmission ($\mu$) & 0.0393 (3.93\%) & 0.0312 (3.12\%) \\
Median Transmission & 0.0392 (3.92\%) & 0.0310 (3.10\%) \\
Standard Deviation ($\sigma$) & 0.0017 & 0.0025 \\
Coefficient of Variation ($\text{CV} = {\sigma}/{\mu}$) & 4.32\% & 8.01\% \\
\bottomrule
\end{tabular}
}
\end{table}

\section{Conclusion}
\label{sec:conclusion}
In this proceeding, we have presented the development, fabrication, and experimental evaluation of a custom $7\times1$ multi-mode photonic lantern (MMPL), designed to serve as a multi-aperture fiber feed for efficient light injection from the MARCOT Pathfinder array at the Calar Alto Observatory. 

Our optical characterization of both bare and connectorized MMPLs demonstrates that the developed lantern achieves exceptional channel-to-channel uniformity, with a coefficient of variation (CV) of less than $10 \%$ across a region of interest spanning 600 to 608 nm. However, the total throughput efficiency of this initial prototype is limited to below $4 \%$. To investigate the cause of this low efficiency, we examined the refractive index profile of the cross-section of the tapered region of the MMPL. We identified a critical waveguiding failure mode resulting from a severe index mismatch between the internal fiber cladding layer and the surrounding glass capillary material. The refractive index of the fiber's inner cladding closely matches that of the capillary, which significantly degrades the index contrast ($\Delta n$) necessary for effective light confinement during thermal tapering. This breakdown in waveguide boundary conditions suppresses total internal reflection (TIR) within the lantern’s longitudinal transition zone, causing light to escape and leading to high radiative losses.

While a throughput efficiency of less than $4 \%$ falls short of the requirements for immediate on-sky deployment, evaluating this specific device provides a valuable diagnostic benchmark for the broader EMARCOT framework. Moving forward, future iterations will focus on structural and material optimizations, such as modifying the numerical aperture (NA) or indices of the capillaries, to prevent cladding-confinement failures. These remediation strategies will directly inform upcoming manufacturing runs, creating a clear pathway toward achieving the high-throughput performance necessary for precision radial-velocity astronomy and high-resolution spectroscopy.

\section{Acknowledgments}
The authors would like to express their sincere gratitude to Dr. Adrian Lorenz, Dr. Tobias Habisreuther, and Dr. Jörg Bierlich, all from IPHT, for fabricating the custom capillary tubes used in this study. We also extend our thanks to Mr. Dennis Plüschke (AIP) for his meticulous execution of the fiber-gluing process. Additionally, we are grateful to Mr. Svend-Marian Bauer (AIP) for designing the 3D-printed protective plate and the milled plastic baseplate, which provided essential mechanical stabilization and support for the photonic lanterns. A.S.N., J.G., and K.M. gratefully acknowledge financial support for this work from the PICS4SENS project, funded by the State of Brandenburg through the Investitionsbank des Landes Brandenburg (ILB), with support from the European Regional Development Fund (ERDF/EFRE), grant number 86000879.

\section{Declaration of AI use}
During the preparation of this work, the authors used Grammarly and Gemini 3.5 Flash to improve the language, readability, and grammatical structure of the manuscript. After using these tools, the authors reviewed, verified, and edited the content as needed, and take full responsibility for the final text of the publication.

\bibliography{report} 
\bibliographystyle{spiebib} 

\end{document}